\documentclass[titlepage,12pt]{article}
\usepackage{amssymb}
\usepackage{amsfonts}
\usepackage{geometry}
\usepackage{mathrsfs}
\usepackage[onehalfspacing]{setspace}
\usepackage{rotating}

\newtheorem{theorem}{Theorem}

\input{tcilatex}
\renewcommand{\mathcal}{\mathscr}

\begin{document}

\title{Doubly stochastic models for replicated spatio-temporal point
processes}
\author{Daniel Gervini \\
%EndAName
Department of Mathematical Sciences\\
University of Wisconsin--Milwaukee}
\maketitle

\begin{abstract}
This paper proposes a log-linear model for the latent intensity functions of
a replicated spatio-temporal point process. By simultaneously fitting
correlated spatial and temporal Karhunen--Lo\`{e}ve expansions, the model
produces spatial and temporal components that are usually easy to interpret
and capture the most important modes of variation and spatio-temporal
correlation of the process. The asymptotic distribution of the estimators is
derived. The finite sample properties are studied by simulations. As an
example of application, we analyze bike usage patterns on the Divvy bike
sharing system of the city of Chicago.

\end{abstract}

\section{Introduction\label{sec:Introd}}

Point processes in time and space have a broad range of applications, in
areas as diverse as neuroscience, ecology, finance, seismology, and many
others. Examples are given in classic texts like Baddeley (2007), Cox and
Isham (1980), Diggle (2013), M\o ller and Waagepetersen (2004) and Streit
(2010). Due to the prevailing types of data and applications, the
point-process literature has mainly focused on single realizations of point
processes, such as the distribution of cells in a single tissue sample
(Diggle et al., 2006). Spatio-temporal processes, in particular, have been
widely studied in the literature (see e.g.~Li and Guan, 2014; Shirota and
Gelfand, 2017; Waagepetersen et al., 2016), but always in the context of
single realizations. An exception has been spike train data, where neural
activity from several patients, or from the same patient under different
trials, is observed (Brown et al., 2004); in such cases we have a replicated
point process, that is, a process observed on different subjects or units.

Among the few papers that have addressed replicated point processes in past
years we can cite Diggle et al.~(1991), Baddeley et al.~(1993), Diggle et
al.~(2000), Bell and Grunwald (2004), Landau et al.~(2004), Wager et
al.~(2004), and Pawlas (2011). Due to their limited amount data, these
papers only proposed estimators for summary statistics of the processes (the
so-called $F$, $G$ and $K$ statistics) rather than the intensity functions
that characterize the processes, which would have been much more informative.

The increased availability of complex data has made replicated point
processes more common in recent years. For example, bike sharing systems are
becoming ubiquitous in large cities around the world (Shaheen et al., 2010).
These systems provide short-term bicycle rental services at unattended
stations distributed within the city. The Divvy system of the city of
Chicago keeps records of every bike trip in the system and makes them
publicly available at the Chicago Data Portal website
(https://data.cityofchicago.org). In this paper (Section \ref{sec:Example})
we will analyze trips that took place between April 1 and November 30 of
2016, since bike usage considerably decreases during the winter. There were
a total of $3,068,211$ bike trips and 458 active bike stations in that
period. Each bike trip can be seen as an observation $(t,\mathbf{s})$ of a
spatio-temporal process, where $t$ is the starting time of the trip and $%
\mathbf{s}$ its destination. We can see the $244$ days between April 1 and
November 30 as the $n$ replications of the spatio-temporal process. We will
focus on trips originating from a single bike station, the one at the
intersection of Wrightwood and Ashland avenues, identified as station 166 in
the system. We chose this station because it has the median total annual
trip count ($4,304$) among the 458 stations in the system. It is of
interest, in this case, to investigate the existence of patterns in the
daily distributions of trip start times and destinations. For example, is
bike demand distributed uniformly during the day, or does it spike at
certain times of the day? And if so, is this pattern similar every day of
the week or is there a difference, for instance, between weekdays and
weekends? Are trip destinations uniformly distributed in the vicinity of the
bike station or do some specific locations attract most trips? And if so,
are these temporal and spatial patterns related? For example, do days with
an increased bike demand on a given time frame (e.g.~early morning) also
show trip destinations concentrated in a specific area (e.g.~downtown)?
Answering these questions is important for an efficient administration of
the system, since understanding the patterns of usage of each station helps
correct the imbalances in bike distribution that inevitably arise in these
systems (Nair and Miller-Hooks, 2011).

Estimating daily spatio-temporal distributions is possible for the methods
proposed in this paper because of the availability of replications, which
allow `borrowing strength'\ across several days. Otherwise, estimation of
daily intensity functions would not be feasible for these data, where some
days only a dozen or so trips take place; such low counts do not allow
accurate estimation of temporal intensity functions, let alone spatial ones,
if each day is estimated separately from the others.

The idea of `borrowing strength'\ across replications underlies most
Functional Data methods (Ramsay and Silverman, 2005). However, Functional
Data Analysis has mostly focused on continuous processes; little work has
been done on discrete point processes so far. There is a link, however,
between discrete and continuous time processes via the underlying intensity
functions, which, although not directly observable, can be seen as
realizations of a latent continuous stochastic process. This relationship
has been exploited by some authors, but the literature in the area is still
scant. We can mention Bouzas et al.~(2006, 2007) and Fern\'{a}ndez-Alcal\'{a}
et al.\ (2012), which have rather limited scopes since they only estimate
the mean of a temporal process, not its variability, and Wu et al.~(2013),
who estimate the mean and the principal components of a temporal process,
but their kernel-based methods are not easy to extend to spatial domains.

The author and his collaborators have recently proposed models for
replicated temporal or spatial processes (Gervini, 2016; Gervini and Khanal,
2019), and for marked point processes with continuous marks (Gervini and
Baur, 2017), but not for jointly spatio-temporal processes. This paper
proposes a log-linear model for the latent intensity process. The model is
based on the Karhunen--Lo\`{e}ve expansion, or principal component
decomposition, of stochastic processes. By fitting correlated temporal\ and
spatial Karhunen--Lo\`{e}ve expansions simultaneously, the model produces
temporal and spatial components that are easy to interpret and capture the
most important modes of variation and spatio-temporal correlation of the
process. Note that this is not simply a matter of fitting separate temporal
and spatial models as in e.g.~Gervini (2016) and Gervini and Khanal (2019)
and then computing the cross-correlations. If this is done, there will
typically appear a `size component' in both models, with a trivially high
cross-correlation because overall count is being explained twice. On the
other hand, components that are important for explaining variability in the
temporal or spatial domains taken separately, may not be optimal for
explaining spatio-temporal cross-correlations; the situation, in this sense,
is similar to that of principal component analysis versus canonical variates
in multivariate analysis (Seber, 2004). Therefore, although there are
similarities with Gervini and Khanal (2019), the joint spatio-temporal
models we propose here are not straightforward extensions of those.

This paper is organized as follows. The new model is presented in Section %
\ref{sec:Model} and its estimation procedure in Section \ref{sec:Estimation}%
. Asymptotic results for statistical inference are derived in Section \ref%
{sec:Asymptotics}, and the finite-sample behavior of the method is studied
by simulation in Section \ref{sec:Simulations}. As an example of
application, the Divvy bike data is analyzed in more detail in Section \ref%
{sec:Example}.

\section{Doubly stochastic spatio-temporal model\label{sec:Model}}

A spatio-tamporal point process $X$ is a random countable set in $\mathcal{S}%
=\mathbb{R}\times \mathbb{R}^{2}$ (M\o ller and Waagepetersen, 2004, ch.~2;
Streit, 2010, ch.~2). A point process is locally finite if $\#(X\cap
B)<\infty $ with probability one for any bounded $B\subseteq \mathcal{S}$.
For a locally finite process we can define the count function $N(B)=\#(X\cap
B)$, which characterizes the distribution of the process. A Poisson process
is a locally finite process for which there exists a locally integrable
function $\lambda :\mathcal{S}\rightarrow \lbrack 0,\infty )$, called the
intensity function, such that \emph{(i)} $N(B)$ has a Poisson distribution
with rate $\int_{B}\lambda (t,\mathbf{s})~dt~d\mathbf{s}$, and \emph{(ii)}
for disjoint sets $B_{1},\ldots ,B_{k}$ the random variables $%
N(B_{1}),\ldots ,N(B_{k})$ are independent. A consequence of \emph{(i)} and 
\emph{(ii)} is that the conditional distribution of the points in $X\cap B$
given $N(B)=m$ is the distribution of $m$ independent and identically
distributed observations with density $\lambda (t,\mathbf{s}%
)/\int_{B}\lambda $.

It follows that for a realization $x=\{(t_{1},\mathbf{s}_{1}),\ldots ,(t_{m},%
\mathbf{s}_{m})\}$ of a Poisson process $X$ on a given bounded region $%
B=B_{t}\times B_{s}$ the density function (in the sense of Proposition 3.1
of M\o ller and Waagepetersen, 2004) is 
\begin{equation}
f(x)=\frac{\exp (-\int_{B}\lambda )}{m!}\prod_{j=1}^{m}\lambda (t_{j},%
\mathbf{s}_{j}).  \label{eq:f_pdf}
\end{equation}

For replicated point processes, a single intensity function $\lambda $
rarely provides an adequate fit for all replications; it is more reasonable
to assume that $\lambda $ itself is random. Such processes are called doubly
stochastic or Cox processes (M\o ller and Waagepetersen, 2004, ch.~5;
Streit, 2010, ch.~8). A doubly stochastic Poisson process is a pair $%
(X,\Lambda )$ where $X|\Lambda =\lambda $ is a Poisson process with
intensity function $\lambda $, and $\Lambda $ is a random function that
takes values on the space $\mathcal{F}$ of non-negative locally integrable
functions on $\mathcal{S}$.

We assume $\Lambda (t,\mathbf{s})$ factorizes as 
\begin{equation}
\Lambda (t,\mathbf{s})=R\Lambda _{t}(t)\Lambda _{s}(\mathbf{s})
\label{eq:lmb_fact}
\end{equation}%
for a temporal process $\Lambda _{t}$, a spatial process $\Lambda _{s}$ and
a random scale factor $R$. Identifiability constraints for this
factorization are discussed below. Factorization (\ref{eq:lmb_fact}) implies
that the overall rate, the distribution of the temporal points and the
distribution of the spatial points are conditionally independent given $%
\Lambda =\lambda $. Therefore the inter-dependence among these three
elements is determined by the dependence structure of $R$, $\Lambda _{t}$
and $\Lambda _{s}$.

The scale factor $R$ and the latent processes $\Lambda _{t}$ and $\Lambda
_{s}$ are non-negative, so for simplicity we will assume they are positive
and model their logarithms: 
\begin{equation}
\log R=\tau +Z,  \label{eq:Log_R}
\end{equation}%
where $Z$ is a zero-mean random variable, 
\begin{equation}
\log \Lambda _{t}(t)=\mu (t)+\sum_{k=1}^{p_{1}}U_{k}\phi _{k}(t)
\label{eq:KL_Lmb_t}
\end{equation}%
and 
\begin{equation}
\log \Lambda _{s}(\mathbf{s})=\nu (\mathbf{s})+\sum_{k=1}^{p_{2}}V_{k}\psi
_{k}(\mathbf{s}),  \label{eq:KL_Lmb_s}
\end{equation}%
where the $\phi _{k}$s and $\psi _{k}$s are orthonormal functions in $%
L^{2}(B_{t})$ and $L^{2}(B_{s})$, respectively, $E(U_{k})=E(V_{k})=0$ for
all $k$, and $\limfunc{cov}(U_{k},U_{k^{\prime }})=\limfunc{cov}%
(V_{k},V_{k^{\prime }})=0$ for all $k\neq k^{\prime }$. The terms in (\ref%
{eq:KL_Lmb_t}) and (\ref{eq:KL_Lmb_s}) are arranged in decreasing order of
variances, $\sigma _{uk}^{2}=\func{var}(U_{k})$ and $\sigma _{vk}^{2}=\func{%
var}(V_{k})$. Note that for any processes $\log \Lambda _{t}\in L^{2}(B_{t})$
with $E(\Vert \log \Lambda _{t}\Vert ^{2})<\infty $ and $\log \Lambda
_{s}\in L^{2}(B_{s})$ with $E(\Vert \log \Lambda _{s}\Vert ^{2})<\infty $,
expansions (\ref{eq:KL_Lmb_t}) and (\ref{eq:KL_Lmb_s}) always hold with
possibly infinite $p_{1}$ and $p_{2}$, and are known as Karhunen--Lo\`{e}ve
expansions (Ash and Gardner, 1975, ch.~1.4). By taking finite $p_{1}$ and $%
p_{2}$ in (\ref{eq:KL_Lmb_t}) and (\ref{eq:KL_Lmb_s}) we do not lose much in
practice, since we are mainly interested in smooth processes where the first
few components dominate.

Factorization (\ref{eq:lmb_fact}) needs some additional constraints for
identifiability. It would seem natural to require that $\Lambda _{t}$ and $%
\Lambda _{s}$ integrate to one, so the overall rate of the process would be $%
R$, and $\Lambda _{t}$ and $\Lambda _{s}$ would be probability density
functions. Unfortunately those constraints are not well adapted to the
log-linear models (\ref{eq:KL_Lmb_t}) and (\ref{eq:KL_Lmb_s}). For
computational simplicity, we will ask instead that $\log \Lambda _{t}$ and $%
\log \Lambda _{s}$ integrate to zero, for which it is sufficient to ask that 
$\mu $, the $\phi _{k}$s, $\nu $ and the $\psi _{k}$s integrate to zero.
These constraints are computationally easy to handle. Under these
conditions, we have 
\[
\log R=\frac{1}{\left\vert B\right\vert }\iint_{B}\log \Lambda (t,\mathbf{s}%
)\ dt\ d\mathbf{s}, 
\]%
\[
\log \Lambda _{t}(t)=\frac{1}{\left\vert B_{s}\right\vert }\int_{B_{s}}\log
\Lambda (t,\mathbf{s})\ d\mathbf{s}-\frac{1}{\left\vert B\right\vert }%
\iint_{B}\log \Lambda (t,\mathbf{s})\ dt\ d\mathbf{s} 
\]%
and 
\[
\log \Lambda _{s}(\mathbf{s})=\frac{1}{\left\vert B_{t}\right\vert }%
\int_{B_{t}}\log \Lambda (t,\mathbf{s})\ dt-\frac{1}{\left\vert B\right\vert 
}\iint_{B}\log \Lambda (t,\mathbf{s})\ dt\ d\mathbf{s}, 
\]%
where $\left\vert \cdot \right\vert $ denotes Lebesgue measure of the
respective sets.

From (\ref{eq:KL_Lmb_t}) and (\ref{eq:KL_Lmb_s}) it follows that the
dependence between $R$, $\Lambda _{t}$ and $\Lambda _{s}$ is determined by
the dependence between $Z$, $\mathbf{U}=(U_{1},\ldots ,U_{p_{1}})^{T}$ and $%
\mathbf{V}=(V_{1},\ldots ,V_{p_{2}})^{T}$. To model this dependence we
collect these random effects into a single vector $\mathbf{W}=(Z,\mathbf{U}%
^{T},\mathbf{V}^{T})^{T}$, which we assume to follow a multivariate normal
distribution with mean zero and covariance matrix 
\[
\mathbf{\Sigma }=\left( 
\begin{array}{ccc}
\sigma _{z}^{2} & \mathbf{\sigma }_{zu}^{T} & \mathbf{\sigma }_{zv}^{T} \\ 
\mathbf{\sigma }_{zu} & \func{diag}(\mathbf{\sigma }_{u}^{2}) & \mathbf{%
\Sigma }_{uv} \\ 
\mathbf{\sigma }_{zv} & \mathbf{\Sigma }_{uv}^{T} & \func{diag}(\mathbf{%
\sigma }_{v}^{2})%
\end{array}%
\right) , 
\]%
where $\mathbf{\sigma }_{u}^{2}=(\sigma _{u1}^{2},\ldots ,\sigma
_{up_{1}}^{2})$, $\mathbf{\sigma }_{v}^{2}=(\sigma _{v1}^{2},\ldots ,\sigma
_{vp_{2}}^{2})$, $\mathbf{\sigma }_{zu}=\func{cov}(Z,\mathbf{U})$, $\mathbf{%
\sigma }_{zv}=\func{cov}(Z,\mathbf{V})$ and $\mathbf{\Sigma }_{uv}=\limfunc{%
cov}(\mathbf{U},\mathbf{V})$. The main parameters of interest here are the
cross-covariances $\mathbf{\sigma }_{zu}$, $\mathbf{\sigma }_{zv}$ and $%
\mathbf{\Sigma }_{uv}$, since they determine the dependence or independence
of the random effects $Z$, $U_{k}$s and $V_{k}$s. In Section \ref%
{sec:Asymptotics} we derive the asymptotic distribution of the estimators of
these parameters with the main goal of obtaining tests and confidence
intervals for inference. Of secondary importance, but still useful, are
confidence intervals for the variances $\sigma _{uk}^{2}$s and $\sigma
_{vk}^{2}$s, since variances that are not significantly different from zero
would indicate that the respective components superfluous.

To facilitate estimation of the functional parameters $\mu $, $\nu $, $\phi
_{k}$s and $\psi _{k}$s, we will use semiparametric basis-function
expansions. As basis functions for the temporal elements we will use $B$%
-splines, and for the spatial elements we will use renormalized Gaussian
radial kernels. But other families could be used, like simplicial bases for
irregular spatial domains; our derivations in this paper are not tied down
to any specific bases. We will call these families $\mathcal{B}_{t}$ and $%
\mathcal{B}_{s}$, respectively. Let $\mathbf{\beta }_{t}(t)$ be the vector
of $q_{1}$ basis functions of $\mathcal{B}_{t}$ and $\mathbf{\beta }_{s}(%
\mathbf{s})$ the vector of $q_{2}$ basis functions of $\mathcal{B}_{s}$.
Then we assume $\mu (t)=\mathbf{c}_{0}^{T}\mathbf{\beta }_{t}(t)$, $\phi
_{k}(t)=\mathbf{c}_{k}^{T}\mathbf{\beta }_{t}(t)$, $\nu (\mathbf{s})=\mathbf{%
d}_{0}^{T}\mathbf{\beta }_{s}(\mathbf{s})$, and $\psi _{k}(\mathbf{s})=%
\mathbf{d}_{k}^{T}\mathbf{\beta }_{s}(\mathbf{s})$. The orthonormality
constraints on the $\phi _{k}$s can be expressed as $\mathbf{c}_{k}^{T}%
\mathbf{J}_{t}\mathbf{c}_{k^{\prime }}=\delta _{kk^{\prime }}$, where $%
\delta _{kk^{\prime }}$ is Kronecker's delta and $\mathbf{J}_{t}=\int_{B_{t}}%
\mathbf{\beta }_{t}(t)\mathbf{\beta }_{t}(t)^{T}dt$, and similarly for the $%
\psi _{k}$s. The zero-integral constraints for $\mu $ and the $\phi _{k}$s
can be expressed as $\mathbf{a}_{t0}^{T}\mathbf{c}_{k}=0$ for $k=0,\ldots
,p_{1}$, where $\mathbf{a}_{t0}=\int_{B_{t}}\mathbf{\beta }_{t}(t)dt$, and
similarly for $\nu $ and the $\psi _{k}$s. For some applications, such as
the bike data mentioned in the Introduction, it is also natural to require
that the temporal intensity functions and their derivatives match at the
endpoints of $B_{t}$. So, if $B_{t}=[t_{l},t_{u}]$, we also have the
constraints $\mu (t_{l})=\mu (t_{u})$, $\mu ^{\prime }(t_{l})=\mu ^{\prime
}(t_{u})$, $\phi _{k}(t_{l})=\phi _{k}(t_{u})$ and $\phi _{k}^{\prime
}(t_{l})=\phi _{k}^{\prime }(t_{u})$ for all $k$, which can be expressed as $%
\mathbf{A}_{P}\mathbf{c}_{k}=\mathbf{0}$ for $k=0,\ldots ,p_{1}$, with $%
\mathbf{A}_{P}=[\mathbf{\beta }_{t}(t_{u})-\mathbf{\beta }_{t}(t_{l}),%
\mathbf{\beta }_{t}^{\prime }(t_{u})-\mathbf{\beta }_{t}^{\prime
}(t_{l})]^{T}$.

\section{Parameter estimation\label{sec:Estimation}}

\subsection{Penalized maximum likelihood estimation}

For simplicity of notation we collect all model parameters into a single
vector 
\begin{equation}
\mathbf{\theta }=(\mathbf{\sigma }_{zu},\mathbf{\sigma }_{zv},\func{vec}%
\mathbf{\Sigma }_{uv},\tau ,\sigma _{z}^{2},\mathbf{c}_{0},\func{vec}\mathbf{%
C},\mathbf{\sigma }_{u}^{2},\mathbf{d}_{0},\func{vec}\mathbf{D},\mathbf{%
\sigma }_{v}^{2}),  \label{eq:theta}
\end{equation}%
where $\mathbf{C}=[\mathbf{c}_{1},\ldots ,\mathbf{c}_{p_{1}}]$ and $\mathbf{D%
}=[\mathbf{d}_{1},\ldots ,\mathbf{d}_{p_{2}}]$. From the distributional
assumptions in Section \ref{sec:Model}, the joint density of $(x,\mathbf{w})$
can be factorized as 
\[
f_{\mathbf{\theta }}(x,\mathbf{w})=f_{\mathbf{\theta }}(x\mid \mathbf{w})f_{%
\mathbf{\theta }}(\mathbf{w}) 
\]%
with $f_{\mathbf{\theta }}(x\mid \mathbf{w})$ as in (\ref{eq:f_pdf}) and $f_{%
\mathbf{\theta }}(\mathbf{w})$ the multivariate normal density. Explicitly, 
\[
f_{\mathbf{\theta }}(x\mid \mathbf{w})=\frac{\exp \{-rI_{t}(\mathbf{u})I_{s}(%
\mathbf{v})\}}{m!}\ r^{m}\prod_{j=1}^{m}\lambda _{t}(t_{j};\mathbf{u}%
)\prod_{j=1}^{m}\lambda _{s}(\mathbf{s}_{j};\mathbf{v}), 
\]%
where $r=\exp (\tau +z)$, $\lambda _{t}(t;\mathbf{u})=\exp \{\mu (t)+\mathbf{%
u}^{T}\mathbf{\phi }(t)\}$, $\lambda _{s}(\mathbf{s};\mathbf{v})=\exp \{\nu (%
\mathbf{s})+\mathbf{v}^{T}\mathbf{\psi }(\mathbf{s})\}$, $I_{t}(\mathbf{u}%
)=\int_{B_{t}}\lambda _{t}(t;\mathbf{u})dt$, and $I_{s}(\mathbf{v}%
)=\int_{B_{s}}\lambda _{s}(\mathbf{s};\mathbf{v})d\mathbf{s}$. The marginal
density for the observable datum $x$ is 
\[
f_{\mathbf{\theta }}(x)=\int f_{\mathbf{\theta }}(x,\mathbf{w})d\mathbf{w,} 
\]%
which has no closed form. We use Laplace's approximation for its evaluation,
as explained in the Supplementary Material.

Given $n$ independent realizations $x_{1},\ldots ,x_{n}$ of the process $X$,
the maximum likelihood estimator of $\mathbf{\theta }$ would be the
maximizer of $\sum_{i=1}^{n}\log f_{\mathbf{\theta }}(x_{i})$. However, when
the basis families $\mathcal{B}_{t}$ and $\mathcal{B}_{s}$ have large
dimensions, it is advisable to regularize the estimators by adding roughness
penalties to the objective function. We then define the penalized
log-likelihood function 
\begin{equation}
\ell _{n}(\mathbf{\theta })=\frac{1}{n}\sum_{i=1}^{n}\log f_{\mathbf{\theta }%
}(x_{i})-\xi _{1}P(\mu )-\xi _{2}\sum_{k=1}^{p_{1}}P(\phi _{k})-\xi
_{3}P(\nu )-\xi _{4}\sum_{k=1}^{p_{2}}P(\psi _{k}),  \label{eq:Pen-log-lik}
\end{equation}%
where the $\xi $s are nonnegative smoothing parameters and the $P(f)$s are
roughness penalty functions. For the temporal functions $\mu $ and $\phi
_{k} $s we use $P(f)=\int (f^{\prime \prime })^{2}$, and for the spatial
functions $\nu $ and $\psi _{k}$s we use $P(f)=\iint \{(\frac{\partial ^{2}f%
}{\partial s_{1}^{2}})^{2}+2(\frac{\partial ^{2}f}{\partial s_{1}\partial
s_{2}})^{2}+(\frac{\partial ^{2}f}{\partial s_{2}^{2}})^{2}\}$. The
estimator of $\mathbf{\theta }$ is then defined as 
\[
\mathbf{\hat{\theta}}=\arg \max_{\mathbf{\theta }\in \Theta }\ell _{n}(%
\mathbf{\theta }), 
\]%
where $\Theta $ is the parameter space that includes all the constraints
discussed in Section \ref{sec:Model}: 
\begin{eqnarray}
\Theta &=&\{\mathbf{\theta }\in \mathbb{R}^{r}:h_{kl}^{C}(\mathbf{\theta }%
)=0,\ k=1,\ldots ,l,\ l=1,\ldots ,p_{1};  \label{eq:Theta_prelim} \\
&&h_{kl}^{D}(\mathbf{\theta })=0,\ k=1,\ldots ,l,\ l=1,\ldots ,p_{2};\ \ 
\mathbf{a}_{t0}^{T}\mathbf{c}_{k}=0,~k=0,\ldots ,p_{1};  \nonumber \\
&&\mathbf{a}_{s0}^{T}\mathbf{d}_{k}=0,~k=0,\ldots ,p_{2};\ \ \mathbf{A}_{P}%
\mathbf{c}_{k}=0,~k=0,\ldots ,p_{1};\ \ \mathbf{\Sigma }>0\},  \nonumber
\end{eqnarray}%
with $r$ the dimension of $\mathbf{\theta }$, $h_{kl}^{C}(\mathbf{\theta })=%
\mathbf{c}_{k}^{T}\mathbf{J}_{t}\mathbf{c}_{l}-\delta _{kl}$, $h_{kl}^{D}(%
\mathbf{\theta })=\mathbf{d}_{k}^{T}\mathbf{J}_{s}\mathbf{d}_{l}-\delta
_{kl} $, and $\mathbf{\Sigma }>0$ denoting that $\mathbf{\Sigma }$ is
symmetric positive definite. The periodicity constraints $\mathbf{A}_{P}%
\mathbf{c}_{k}=0$ need not be present in every situation, but for generality
we include them in all our derivations; the results below are still valid if
these constraints are simply deleted.

Once $\mathbf{\hat{\theta}}$ has been obtained, individual predictors of the
latent random effects $\mathbf{w}$ can be obtained as $\mathbf{\hat{w}}%
_{i}=E_{\mathbf{\hat{\theta}}}(\mathbf{w}\mid x_{i})$. The estimating
equations for $\mathbf{\hat{\theta}}$ and an EM algorithm (Dempster et al.,
1977) for its computation are derived in the Supplementary Material.
Programs implementing these algorithms are available on the author's website.

\subsection{Choice of meta-parameters}

The proposed model has a number of tuning parameters that have to be chosen
by the user: \emph{(i)} the number of functional components $p_{1}$ and $%
p_{2}$, \emph{(ii)} the basis families $\mathcal{B}_{t}$ and $\mathcal{B}%
_{s} $ and in particular their dimensions $q_{1}$ and $q_{2}$, and \emph{%
(iii)} the smoothing parameters $\xi $s. Regarding \emph{(ii)} we can say
that the overall dimensions $q_{1}$ and $q_{2}$ of the basis families are
more relevant parameters than the specifics of the basis functions such as
e.g.~the precise knot placement or the degree of the polynomials used for a
spline family. The dimensions of $\mathcal{B}_{t}$ and $\mathcal{B}_{s}$
should be chosen relatively large in order to avoid bias; the variability of
the estimators will be taken care of by the $\xi $s. As noted by Ruppert
(2002, sec.~3), although optimal $q_{1}$ and $q_{2}$ could be chosen by
cross-validation (Hastie et al., 2009, ch.~7), there is little change in
goodness of fit after a minimum dimension has been reached, and the fit is
essentially determined by the smoothing parameters thereafter.

The choice of the smoothing parameters $\xi $s is then more important. It
can be done objectively by cross-validation. Leave-one-out cross-validation
finds $\xi $s that maximize 
\begin{equation}
\mathrm{CV}(\xi _{1},\xi _{2},\xi _{3},\xi _{4})=\sum_{i=1}^{n}\log f_{%
\mathbf{\hat{\theta}}^{[-i]}}(x_{i}),  \label{eq:CV}
\end{equation}%
where $\mathbf{\hat{\theta}}^{[-i]}$ denotes the estimator obtained without
observation $x_{i}$. A faster alternative is to use $k$-fold
cross-validation, where the data is split into $k$ subsets that are
alternatively used as test data; $k=5$ is a common choice. Full
four-dimensional optimization of (\ref{eq:CV}) is too time consuming even
for five-fold cross-validation; a workable alternative is sequential
optimization, where each $\xi _{j}$ in turn is optimized on a grid while the
other $\xi $s are kept fixed at an initial value chosen by the user. A
faster but subjective alternative is to choose the $\xi $s by visual
inspection. Plots of the means and the components for different $\xi $s can
be inspected to see how new features of the curves appear or disappear as $%
\xi $ decreases or increases; the $\xi $s that produce curves with
well-defined but not too irregular features can then be chosen. Curve shapes
change smoothly with $\xi $, so there is usually a relatively broad range of 
$\xi $s that will produce comparable and reasonable results; there is no
need to select the exact optimal $\xi $.

The choice of the number of components $p_{1}$ and $p_{2}$ can also be done
either objectively or subjectively, the former by cross-validation or
testing, the latter by taking into account the relative contributions of the
new components on the total variances, $\sigma _{up_{1}}^{2}/(\sigma
_{u1}^{2}+\cdots +\sigma _{up_{1}}^{2})$ and $\sigma _{vp_{2}}^{2}/(\sigma
_{v1}^{2}+\cdots +\sigma _{vp_{2}}^{2})$.

\section{Asymptotics\label{sec:Asymptotics}}

The asymptotic behavior of $\mathbf{\hat{\theta}}$ as $n\rightarrow \infty $
can be studied via empirical-process techniques (Pollard, 1984; Van der
Vaart, 2000), since (\ref{eq:Pen-log-lik}) is the average of independent
identically distributed functions plus non-random roughness penalties, as in
e.g.~Knight and Fu (2000). We will develop here a `parametric' asymptotics
where the dimensions $q_{1}$ and $q_{2}$ of the basis families $\mathcal{B}%
_{t}$ and $\mathcal{B}_{s}$ are held fixed and the true functional
parameters are assumed to belong to $\mathcal{B}_{t}$ and $\mathcal{B}_{s}$.
This assumption, in practice, is not very unrealistic as long as $q_{1}$ and 
$q_{2}$ are reasonably large. Other authors have followed this `parametric'
asymptotic approach in similar semiparametric contexts (e.g.~Yu and Ruppert,
2002, and Xun et al., 2013).

The first result of this section, Theorem \ref{thm:Constcy}, establishes
consistency of the estimator $\mathbf{\hat{\theta}}$. The proof is given in
the Supplementary Material. For uniqueness of the true parameters, the
indeterminate signs of the $\phi _{k}$s and $\psi _{k}$s require special
handling; we also need to assume that the components have multiplicity one.
Our modified parameter space, then, will be 
\begin{eqnarray}
\Theta &=&\{\mathbf{\theta }\in \mathbb{R}^{r}:h_{kl}^{C}(\mathbf{\theta }%
)=0,\ k=1,\ldots ,l,\ l=1,\ldots ,p_{1};  \label{eq:Theta_final} \\
&&h_{kl}^{D}(\mathbf{\theta })=0,\ k=1,\ldots ,l,\ l=1,\ldots ,p_{2};\ \ 
\mathbf{a}_{t0}^{T}\mathbf{c}_{k}=0,\ k=0,\ldots ,p_{1};  \nonumber \\
&&\mathbf{a}_{s0}^{T}\mathbf{d}_{k}=0,\ k=0,\ldots ,p_{2};\ \ \mathbf{A}_{P}%
\mathbf{c}_{k}=0,\ k=0,\ldots ,p_{1};\ \   \nonumber \\
&&\mathbf{\Sigma }>0;\ \ \sigma _{u1}>\cdots >\sigma _{up_{1}}>0;\ \ \sigma
_{v1}>\cdots >\sigma _{vp_{2}}>0;  \nonumber \\
&&c_{k1}\geq 0,\ k=1,\ldots ,p_{1};\ \ d_{k1}\geq 0,\ k=1,\ldots ,p_{2}\}. 
\nonumber
\end{eqnarray}%
We make the following assumptions:

\begin{description}
\item[A1] The signs of the $\hat{\phi}_{k}$s and $\hat{\psi}_{k}$s are
specified so that the first non-zero basis coefficient of each $\hat{\phi}%
_{k}$ and $\hat{\psi}_{k}$ is positive (then $\mathbf{\hat{\theta}}\in
\Theta $ for $\Theta $ defined in (\ref{eq:Theta_final}).)

\item[A2] The true functional parameters $\mu _{0}$, $\nu _{0}$, $\phi _{k0}$%
s and $\psi _{k0}$s of models (\ref{eq:KL_Lmb_t}) and (\ref{eq:KL_Lmb_s})
belong to the functional spaces $\mathcal{B}_{t}$ and $\mathcal{B}_{s}$, and
their basis coefficients $c_{k1,0}$ and $d_{k1,0}$ are not zero. The signs
of $\phi _{k0}$ and $\psi _{k0}$ are then chosen so that $c_{k1,0}>0$ and $%
d_{k1,0}>0$; therefore there is a unique $\mathbf{\theta }_{0}$ in $\Theta $
such that $f_{\mathbf{\theta }_{0}}(x)$ is the true density of the data.

\item[A3] $\mathbf{\xi }_{n}\rightarrow \mathbf{0}$ as $n\rightarrow \infty $%
, where $\mathbf{\xi }_{n}=(\xi _{1n},\xi _{2n},\xi _{3n},\xi _{4n})^{T}$ is
the vector of smoothing parameters in (\ref{eq:Pen-log-lik}).
\end{description}

The requirement, in assumption A2, that the first basis coefficients $%
c_{k1,0}$ and $d_{k1,0}$ of each $\phi _{k0}$ and $\psi _{k0}$ be non-zero
is somewhat artificial: although the $\phi _{k0}$s and $\psi _{k0}$s must
have at least one non-zero basis coefficient, it need not be the first one.
However, a condition like this is necessary to uniquely identify a\ `true'
parameter $\mathbf{\theta }_{0}$, which would otherwise be unidentifiable
due to sign ambiguity, and that condition has to be consistent with a
sign-specification rule that can be used for the estimators, such as the one
in assumption A1.

\begin{theorem}
\label{thm:Constcy}Under assumptions A1--A3, $\mathbf{\hat{\theta}}%
%TCIMACRO{\TeXButton{TeX field}{\overset{P}{\rightarrow}}}%
%BeginExpansion
\overset{P}{\rightarrow}%
%EndExpansion
\mathbf{\theta }_{0}$ as $n\rightarrow \infty $.
\end{theorem}

To establish asymptotic normality of the estimators we will use the results
of Geyer (1994), which make use of the tangent cone of the parameter space.
The definition and properties of tangent cones can be found in Rockafellar
and Wets (1998, ch.~6). From Theorem 6.31 of Rockafellar and Wets (1998),
the tangent cone of $\Theta $ at $\mathbf{\theta }_{0}$ is 
\begin{eqnarray*}
\mathcal{T}_{0} &=&\{\mathbf{\delta }\in \mathbb{R}^{r}:\nabla h_{kl}^{C}(%
\mathbf{\theta }_{0})^{T}\mathbf{\delta }=0,\ k=1,\ldots ,l,\ l=1,\ldots
,p_{1}; \\
&&\nabla h_{kl}^{D}(\mathbf{\theta }_{0})^{T}\mathbf{\delta }=0,\ k=1,\ldots
,l,\ l=1,\ldots ,p_{2};\ \ \mathbf{a}_{t0}^{T}\mathbf{K}_{\mathbf{c}_{k}}%
\mathbf{\delta }=0,\ k=0,\ldots ,p_{1}; \\
&&\mathbf{a}_{s0}^{T}\mathbf{K}_{\mathbf{d}_{k}}\mathbf{\delta }=0,\
k=0,\ldots ,p_{2};\ \ \mathbf{A}_{P}\mathbf{K}_{\mathbf{c}_{k}}\mathbf{%
\delta }=0,\ k=0,\ldots ,p_{1}\},
\end{eqnarray*}%
where $\mathbf{K}_{\mathbf{d}_{k}}$ and $\mathbf{K}_{\mathbf{c}_{k}}$ are
the `extraction' matrices such that $\mathbf{d}_{k}=\mathbf{K}_{\mathbf{d}%
_{k}}\mathbf{\theta }$ and $\mathbf{c}_{k}=\mathbf{K}_{\mathbf{c}_{k}}%
\mathbf{\theta }$. The explicit forms of $\nabla h_{kl}^{C}(\mathbf{\theta }%
) $ and $\nabla h_{kl}^{D}(\mathbf{\theta })$ are derived in the
Supplementary Material. Let $\mathbf{A}$ be the $r_{1}\times r$ matrix with
rows $\nabla h_{kl}^{C}(\mathbf{\theta }_{0})^{T}$, $\nabla h_{kl}^{D}(%
\mathbf{\theta }_{0})^{T}$, $\mathbf{a}_{t0}^{T}\mathbf{K}_{\mathbf{c}_{k}}$%
, $\mathbf{a}_{s0}^{T}\mathbf{K}_{\mathbf{d}_{k}}$ and $\mathbf{A}_{P}%
\mathbf{K}_{\mathbf{c}_{k}}$, and let $\mathbf{B}$ be an orthogonal
complement of $\mathbf{A}$, that is, an $(r-r_{1})\times r$ matrix such that 
$\mathbf{AB}^{T}=\mathbf{O}$.

The next theorem gives the asymptotic distribution of $\mathbf{\hat{\theta}}$%
. In addition to $\mathbf{B}$ above, it uses Fisher's information matrix, 
\begin{eqnarray*}
\mathbf{F}_{0} &=&E_{\mathbf{\theta }_{0}}\{\nabla \log f_{\mathbf{\theta }%
_{0}}(X)\nabla \log f_{\mathbf{\theta }_{0}}(X)^{T}\} \\
&=&-E_{\mathbf{\theta }_{0}}\{\nabla ^{2}\log f_{\mathbf{\theta }_{0}}(X)\},
\end{eqnarray*}%
where $\nabla $ and $\nabla ^{2}$ are taken with respect to the parameter $%
\mathbf{\theta }$, and $\mathsf{D}\mathbf{P}(\mathbf{\theta })$, the
Jacobian matrix of the smoothness penalty vector $\mathbf{P}(\mathbf{\theta }%
)=(P(\mu ),\sum_{k=1}^{p_{1}}P(\phi _{k}),P(\nu ),\sum_{k=1}^{p_{2}}P(\psi
_{k}))^{T}$. Explicit expressions for these derivatives are given in the
Supplementary Material. We also need an additional assumption:

\begin{description}
\item[A4] $\sqrt{n}\mathbf{\xi }_{n}\rightarrow \mathbf{\kappa }$ as $%
n\rightarrow \infty $, for a finite $\mathbf{\kappa }$.
\end{description}

\begin{theorem}
\label{thm:Asymp}Under assumptions A1--A4, $\sqrt{n}(\mathbf{\hat{\theta}}-%
\mathbf{\theta }_{0})%
%TCIMACRO{\TeXButton{TeX field}{\overset{D}{\rightarrow}}}%
%BeginExpansion
\overset{D}{\rightarrow}%
%EndExpansion
\mathrm{N}(\mathbf{-V}\mathsf{D}\mathbf{P}(\mathbf{\theta }_{0})^{T}\mathbf{%
\kappa },\mathbf{V})$ as $n\rightarrow \infty $, with $\mathbf{V}=\mathbf{B}%
^{T}(\mathbf{BF}_{0}\mathbf{B}^{T})^{-1}\mathbf{B}$.
\end{theorem}

Fisher's information matrix $\mathbf{F}_{0}$ can be estimated by 
\begin{equation}
\mathbf{\hat{F}}_{0}=\frac{1}{n}\sum_{i=1}^{n}\nabla \log f_{\mathbf{\hat{%
\theta}}}(x_{i})\nabla \log f_{\mathbf{\hat{\theta}}}(x_{i})^{T}
\label{eq:F_hat}
\end{equation}%
and $\mathbf{V}$ in Theorem \ref{thm:Asymp} by $\mathbf{\hat{V}}=\mathbf{B}%
^{T}(\mathbf{B\hat{F}}_{0}\mathbf{B}^{T})^{-1}\mathbf{B}$. Due to the high
dimensionality of $\mathbf{\theta }$, $\mathbf{\hat{F}}_{0}$ is often
singular or nearly singular for small sample sizes, leading to unstable
values of $\mathbf{\hat{V}}$. in such cases, a practical alternative is to
treat the functional parameters $\mu $, $\nu $, $\phi _{k}$s and $\psi _{k}$%
s as if they were fixed and known, reducing $\mathbf{\theta }$ to a more
manageable $\mathbf{\tilde{\theta}}=(\mathbf{\sigma }_{zu},\mathbf{\sigma }%
_{zv},\func{vec}\mathbf{\Sigma }_{uv},\tau ,\sigma _{z}^{2},\mathbf{\sigma }%
_{u}^{2},\mathbf{\sigma }_{v}^{2})$. Fisher's information matrix for $%
\mathbf{\tilde{\theta}}$, $\mathbf{\tilde{F}}_{0}$, is usually
low-dimensional enough that it can be accurately estimated by the
corresponding version of (\ref{eq:F_hat}), $\widehat{\mathbf{\tilde{F}}}_{0}$%
, even for relatively small sample sizes. Note that since $\mathbf{\tilde{%
\theta}}$ is not subject to equality constraints or smoothness penalties,
the asymptotic distribution of $\sqrt{n}(\widehat{\mathbf{\tilde{\theta}}}-%
\mathbf{\tilde{\theta}}_{0})$ is simply $\mathrm{N}(\mathbf{0},\mathbf{%
\tilde{F}}_{0}^{-1})$, the standard maximum likelihood asymptotics. The
functional parameters $\mu $, $\nu $, $\phi _{k}$s and $\psi _{k}$s still
need to be estimated, of course, since they must be plugged into $\widehat{%
\mathbf{\tilde{F}}}_{0}$. This `reduced' or `marginal' asymptotics produces
accurate variance estimators even for small sample sizes, as shown by
simulation in Section \ref{sec:Simulations}.

\section{Simulations\label{sec:Simulations}}

To assess the finite-sample behavior of the estimators, we generated data
from model (\ref{eq:Log_R})-(\ref{eq:KL_Lmb_t})-(\ref{eq:KL_Lmb_s}) with $%
p_{1}=p_{2}=2$. We took the interval $B_{t}=[0,1]$ as temporal domain, and
functional parameters $\mu (t)=\sin \pi t-c_{1}$, $\phi _{1}(t)=(\sin \pi
t-c_{1})/c_{2}$ and $\phi _{2}(t)=\sqrt{2}\sin 2\pi t$, where $c_{1}$ and $%
c_{2}$ are standardizing constants. As spatial domain we took the rectangle $%
B_{s}=[0,1]\times \lbrack 0,1]$, and functional parameters $\nu
(s_{1},s_{2})=-(s_{1}-.5)^{2}-(s_{2}-.5)^{2}-c_{3}$, $\psi
_{1}(s_{1},s_{2})=\{\sin (\pi s_{1})\sin (\pi s_{2})-c_{4}\}/c_{5}$ and $%
\psi _{2}(s_{1},s_{2})=2\sin (2\pi s_{1})\sin (2\pi s_{2})$, where $c_{3}$, $%
c_{4}$ and $c_{5}$ are standardizing constants. For $\tau $ we used two
different values, $\tau =\log 10$ and $\tau =\log 30$; the lower $\tau $
generates sparse data where the individual intensity functions cannot be
estimated by individual smoothing.

The variances were taken as $\sigma _{u1}^{2}=.3^{2}\times .7$, $\sigma
_{u2}^{2}=.3^{2}\times .3$, $\sigma _{v1}^{2}=.7^{2}\times .7$ and $\sigma
_{v2}^{2}=.7^{2}\times .3$. The cross-covariance parameters were set as $%
\mathbf{\sigma }_{zu}=\mathbf{0}$, $\mathbf{\sigma }_{zv}=\mathbf{0}$ and $%
\mathbf{\Sigma }_{uv}$ a diagonal matrix with elements $\mathbf{\Sigma }%
_{uv,11}=.7\sigma _{u1}\sigma _{v1}$ and $\mathbf{\Sigma }_{uv,22}=.7\sigma
_{u2}\sigma _{v2}$, so $U_{1}$ and $U_{2}$ were correlated with $V_{1}$ and $%
V_{2}$, respectively. We considered four sample sizes $n$: $50$, $100$, $200$
and $400$. Each scenario was replicated 500 times.

For estimation we used cubic $B$-splines with ten equally spaced knots for
the temporal functions, and normalized Gaussian kernels with 25 uniformly
spaced knots for the spatial functions. This gives dimensions $q_{1}=14$ and 
$q_{2}=25$, respectively. As smoothing parameters we took all $\xi $s equal
to $10^{-5}$.

%TCIMACRO{\TeXButton{B}{\begin{table}[tbp] \centering}}%
%BeginExpansion
\begin{table}[tbp] \centering%
%EndExpansion

\begin{tabular}{crcccrrcccc}
\hline
&  & \multicolumn{9}{c}{$\tau $} \\ \cline{3-11}
&  & \multicolumn{4}{c}{$\log 10$} &  & \multicolumn{4}{c}{$\log 30$} \\ 
\cline{3-6}\cline{8-11}
&  & \multicolumn{4}{c}{$n$} &  & \multicolumn{4}{c}{$n$} \\ 
\cline{3-6}\cline{8-11}
Parameter & \multicolumn{1}{c}{} & $50$ & $100$ & $200$ & $400$ & 
\multicolumn{1}{c}{} & $50$ & $100$ & $200$ & $400$ \\ \hline
&  & \multicolumn{1}{r}{} & \multicolumn{1}{r}{} & \multicolumn{1}{r}{} &  & 
& \multicolumn{1}{r}{} & \multicolumn{1}{r}{} & \multicolumn{1}{r}{} &  \\ 
$\sigma _{zu,1}$ &  & \multicolumn{1}{r}{$.024$} & \multicolumn{1}{r}{$.019$}
& \multicolumn{1}{r}{$.016$} & $.013$ &  & \multicolumn{1}{r}{$.015$} & 
\multicolumn{1}{r}{$.013$} & \multicolumn{1}{r}{$.012$} & $.011$ \\ 
$\sigma _{zu,2}$ &  & \multicolumn{1}{r}{$.016$} & \multicolumn{1}{r}{$.011$}
& \multicolumn{1}{r}{$.009$} & $.007$ &  & \multicolumn{1}{r}{$.012$} & 
\multicolumn{1}{r}{$.008$} & \multicolumn{1}{r}{$.006$} & $.004$ \\ 
$\sigma _{zv,1}$ &  & \multicolumn{1}{r}{$.044$} & \multicolumn{1}{r}{$.043$}
& \multicolumn{1}{r}{$.037$} & $.035$ &  & \multicolumn{1}{r}{$.039$} & 
\multicolumn{1}{r}{$.034$} & \multicolumn{1}{r}{$.031$} & $.030$ \\ 
$\sigma _{zv,2}$ &  & \multicolumn{1}{r}{$.030$} & \multicolumn{1}{r}{$.019$}
& \multicolumn{1}{r}{$.013$} & $.008$ &  & \multicolumn{1}{r}{$.022$} & 
\multicolumn{1}{r}{$.014$} & \multicolumn{1}{r}{$.009$} & $.007$ \\ 
$\Sigma _{uv,11}$ &  & \multicolumn{1}{r}{$.043$} & \multicolumn{1}{r}{$.031$%
} & \multicolumn{1}{r}{$.017$} & $.014$ &  & \multicolumn{1}{r}{$.032$} & 
\multicolumn{1}{r}{$.021$} & \multicolumn{1}{r}{$.015$} & $.010$ \\ 
$\Sigma _{uv,21}$ &  & \multicolumn{1}{r}{$.040$} & \multicolumn{1}{r}{$.031$%
} & \multicolumn{1}{r}{$.018$} & $.015$ &  & \multicolumn{1}{r}{$.024$} & 
\multicolumn{1}{r}{$.017$} & \multicolumn{1}{r}{$.011$} & $.008$ \\ 
$\Sigma _{uv,12}$ &  & \multicolumn{1}{r}{$.031$} & \multicolumn{1}{r}{$.019$%
} & \multicolumn{1}{r}{$.012$} & $.008$ &  & \multicolumn{1}{r}{$.014$} & 
\multicolumn{1}{r}{$.013$} & \multicolumn{1}{r}{$.008$} & $.006$ \\ 
$\Sigma _{uv,22}$ &  & \multicolumn{1}{r}{$.026$} & \multicolumn{1}{r}{$.015$%
} & \multicolumn{1}{r}{$.011$} & $.007$ &  & \multicolumn{1}{r}{$.013$} & 
\multicolumn{1}{r}{$.008$} & \multicolumn{1}{r}{$.007$} & $.004$ \\ 
&  & \multicolumn{1}{r}{} & \multicolumn{1}{r}{} & \multicolumn{1}{r}{} &  & 
& \multicolumn{1}{r}{} & \multicolumn{1}{r}{} & \multicolumn{1}{r}{} &  \\ 
$\tau $ &  & \multicolumn{1}{r}{$.089$} & \multicolumn{1}{r}{$.092$} & 
\multicolumn{1}{r}{$.101$} & $.103$ &  & \multicolumn{1}{r}{$.073$} & 
\multicolumn{1}{r}{$.066$} & \multicolumn{1}{r}{$.065$} & $.065$ \\ 
$\mu $ &  & \multicolumn{1}{r}{$.145$} & \multicolumn{1}{r}{$.107$} & 
\multicolumn{1}{r}{$.074$} & $.062$ &  & \multicolumn{1}{r}{$.117$} & 
\multicolumn{1}{r}{$.084$} & \multicolumn{1}{r}{$.072$} & $.058$ \\ 
$\phi _{1}$ &  & \multicolumn{1}{r}{$.551$} & \multicolumn{1}{r}{$.425$} & 
\multicolumn{1}{r}{$.243$} & $.180$ &  & \multicolumn{1}{r}{$.365$} & 
\multicolumn{1}{r}{$.249$} & \multicolumn{1}{r}{$.157$} & $.116$ \\ 
$\phi _{2}$ &  & \multicolumn{1}{r}{$.724$} & \multicolumn{1}{r}{$.541$} & 
\multicolumn{1}{r}{$.395$} & $.349$ &  & \multicolumn{1}{r}{$.451$} & 
\multicolumn{1}{r}{$.296$} & \multicolumn{1}{r}{$.199$} & $.156$ \\ 
$\nu $ &  & \multicolumn{1}{r}{$.291$} & \multicolumn{1}{r}{$.244$} & 
\multicolumn{1}{r}{$.203$} & $.188$ &  & \multicolumn{1}{r}{$.256$} & 
\multicolumn{1}{r}{$.215$} & \multicolumn{1}{r}{$.202$} & $.178$ \\ 
$\psi _{1}$ &  & \multicolumn{1}{r}{$.397$} & \multicolumn{1}{r}{$.295$} & 
\multicolumn{1}{r}{$.217$} & $.165$ &  & \multicolumn{1}{r}{$.274$} & 
\multicolumn{1}{r}{$.213$} & \multicolumn{1}{r}{$.173$} & $.150$ \\ 
$\psi _{2}$ &  & \multicolumn{1}{r}{$.582$} & \multicolumn{1}{r}{$.424$} & 
\multicolumn{1}{r}{$.315$} & $.249$ &  & \multicolumn{1}{r}{$.372$} & 
\multicolumn{1}{r}{$.267$} & \multicolumn{1}{r}{$.204$} & $.171$ \\ 
\multicolumn{1}{r}{} &  &  &  &  &  &  &  &  &  &  \\ 
$\sigma _{z}$ &  & \multicolumn{1}{r}{$.068$} & \multicolumn{1}{r}{$.049$} & 
\multicolumn{1}{r}{$.037$} & $.027$ &  & \multicolumn{1}{r}{$.051$} & 
\multicolumn{1}{r}{$.030$} & \multicolumn{1}{r}{$.022$} & $.018$ \\ 
$\sigma _{u1}$ &  & \multicolumn{1}{r}{$.045$} & \multicolumn{1}{r}{$.034$}
& \multicolumn{1}{r}{$.021$} & $.018$ &  & \multicolumn{1}{r}{$.036$} & 
\multicolumn{1}{r}{$.024$} & \multicolumn{1}{r}{$.017$} & $.012$ \\ 
$\sigma _{u2}$ &  & \multicolumn{1}{r}{$.040$} & \multicolumn{1}{r}{$.032$}
& \multicolumn{1}{r}{$.026$} & $.022$ &  & \multicolumn{1}{r}{$.030$} & 
\multicolumn{1}{r}{$.021$} & \multicolumn{1}{r}{$.016$} & $.011$ \\ 
$\sigma _{v1}$ &  & \multicolumn{1}{r}{$.060$} & \multicolumn{1}{r}{$.057$}
& \multicolumn{1}{r}{$.050$} & $.040$ &  & \multicolumn{1}{r}{$.062$} & 
\multicolumn{1}{r}{$.047$} & \multicolumn{1}{r}{$.034$} & $.031$ \\ 
$\sigma _{v2}$ &  & \multicolumn{1}{r}{$.047$} & \multicolumn{1}{r}{$.028$}
& \multicolumn{1}{r}{$.021$} & $.016$ &  & \multicolumn{1}{r}{$.037$} & 
\multicolumn{1}{r}{$.027$} & \multicolumn{1}{r}{$.022$} & $.017$ \\ \hline
\end{tabular}

\caption{Simulation Results. Root mean squared errors of parameter
estimators.}\label{tab:Sim_errors_params}%
%TCIMACRO{\TeXButton{E}{\end{table}}}%
%BeginExpansion
\end{table}%
%EndExpansion

%TCIMACRO{\TeXButton{B}{\begin{table}[tbp] \centering}}%
%BeginExpansion
\begin{table}[tbp] \centering%
%EndExpansion

\begin{tabular}{crcccrrcccc}
\hline
&  & \multicolumn{9}{c}{$\tau $} \\ \cline{3-11}
&  & \multicolumn{4}{c}{$\log 10$} &  & \multicolumn{4}{c}{$\log 30$} \\ 
\cline{3-6}\cline{8-11}
&  & \multicolumn{4}{c}{$n$} &  & \multicolumn{4}{c}{$n$} \\ 
\cline{3-6}\cline{8-11}
Variable & \multicolumn{1}{c}{} & $50$ & $100$ & $200$ & $400$ & 
\multicolumn{1}{c}{} & $50$ & $100$ & $200$ & $400$ \\ \hline
\multicolumn{1}{r}{} &  &  &  &  &  &  &  &  &  &  \\ 
$Z$ &  & \multicolumn{1}{r}{$.244$} & \multicolumn{1}{r}{$.242$} & 
\multicolumn{1}{r}{$.232$} & $.230$ &  & \multicolumn{1}{r}{$.179$} & 
\multicolumn{1}{r}{$.174$} & \multicolumn{1}{r}{$.174$} & \multicolumn{1}{r}{%
$.170$} \\ 
$U_{1}$ &  & \multicolumn{1}{r}{$.200$} & \multicolumn{1}{r}{$.180$} & 
\multicolumn{1}{r}{$.171$} & $.167$ &  & \multicolumn{1}{r}{$.158$} & 
\multicolumn{1}{r}{$.144$} & \multicolumn{1}{r}{$.138$} & \multicolumn{1}{r}{%
$.135$} \\ 
$U_{2}$ &  & \multicolumn{1}{r}{$.169$} & \multicolumn{1}{r}{$.157$} & 
\multicolumn{1}{r}{$.148$} & $.144$ &  & \multicolumn{1}{r}{$.132$} & 
\multicolumn{1}{r}{$.121$} & \multicolumn{1}{r}{$.116$} & \multicolumn{1}{r}{%
$.114$} \\ 
$V_{1}$ &  & \multicolumn{1}{r}{$.298$} & \multicolumn{1}{r}{$.282$} & 
\multicolumn{1}{r}{$.271$} & $.265$ &  & \multicolumn{1}{r}{$.239$} & 
\multicolumn{1}{r}{$.215$} & \multicolumn{1}{r}{$.211$} & \multicolumn{1}{r}{%
$.196$} \\ 
$V_{2}$ &  & \multicolumn{1}{r}{$.274$} & \multicolumn{1}{r}{$.247$} & 
\multicolumn{1}{r}{$.237$} & $.230$ &  & \multicolumn{1}{r}{$.189$} & 
\multicolumn{1}{r}{$.173$} & \multicolumn{1}{r}{$.162$} & \multicolumn{1}{r}{%
$.157$} \\ \hline
\end{tabular}

\caption{Simulation Results. Root mean squared errors of random-effect
estimators.}\label{tab:Sim_errors_effects}%
%TCIMACRO{\TeXButton{E}{\end{table}}}%
%BeginExpansion
\end{table}%
%EndExpansion

As measure of estimation error we use the root mean squared error. For
scalar parameters, e.g.~$\tau $, we employ the usual definition, $\{E(\hat{%
\tau}-\tau )^{2}\}^{1/2}$. For functional parameters, e.g.~$\mu (t)$, the
root mean squared error is defined in terms of the $L^{2}$-norm as $%
\{E(\Vert \hat{\mu}-\mu \Vert ^{2})\}^{1/2}$. For random-effect estimators,
e.g.~the $\hat{u}_{i1}$s, we define it as $\{E\sum_{i=1}^{n}(\hat{u}%
_{i1}-u_{i1})^{2}/n\}^{1/2}$. The signs of the $\hat{\phi}_{k}$s and the $%
\hat{\psi}_{k}$s, which in principle are indeterminate, are chosen as the
signs of the inner products $\langle \hat{\phi}_{k},\phi _{k}\rangle $ and $%
\langle \hat{\psi}_{k},\psi _{k}\rangle $, respectively; the signs of the $%
\hat{u}_{ik}$s and $\hat{v}_{ik}$s, and of the elements of $\mathbf{\hat{%
\sigma}}_{zu}$, $\mathbf{\hat{\sigma}}_{zv}$ and $\mathbf{\hat{\Sigma}}_{uv}$%
, are changed accordingly.

Table \ref{tab:Sim_errors_params} shows that, as expected, the estimation
errors decrease as $n$ increases, and also decrease as the baseline rate,
determined by $\tau $, increases. But even in the sparse situation $\tau
=\log 10$ we see that the functional parameters are accurately estimated,
which shows the advantages of `borrowing strength'\ across replications.
Somewhat unusual is the case of $\hat{\tau}$, whose estimation errors do not
decrease as functions of $n$ as fast as they do for the other parameters. A
more in-depth analysis reveals that this is due to bias. Nevertheless, $\tau 
$ is not a very important parameter for inferential purposes; more important
are the cross-covariance parameters and the functional components, and they
are accurately estimated.

Table \ref{tab:Sim_errors_effects} shows that the estimation errors of the
random-effect estimators also decrease as $n$ increases, but $\tau $, which
determines the number of observations per individual, has a larger impact
here than $n$ does. The reason is that each random-effect estimator can only
be computed from the observations available for each individual; `borrowing
strength'\ across replications is not possible for the random effects.

%TCIMACRO{\TeXButton{B}{\begin{table}[tbp] \centering}}%
%BeginExpansion
\begin{table}[tbp] \centering%
%EndExpansion

%TCIMACRO{\TeXButton{Begin sideways}{\begin{sideways}}}%
%BeginExpansion
\begin{sideways}%
%EndExpansion

\begin{tabular}{ccccccccccccccccc}
\hline
&  & \multicolumn{15}{c}{$n$} \\ \cline{3-17}
&  & \multicolumn{3}{c}{$50$} &  & \multicolumn{3}{c}{$100$} &  & 
\multicolumn{3}{c}{$200$} &  & \multicolumn{3}{c}{$400$} \\ 
\cline{3-5}\cline{7-9}\cline{11-13}\cline{15-17}
Parameter &  & True & Mean & Sd &  & True & Mean & Sd &  & True & Mean & Sd
&  & True & Mean & Sd \\ \hline
$\func{sd}(\hat{\sigma}_{zu,1})$ &  & $.211$ & $.355$ & $.082$ &  & $.141$ & 
$.206$ & $.038$ &  & $.104$ & $.128$ & $.015$ &  & $.072$ & $.084$ & $.007$
\\ 
$\func{sd}(\hat{\sigma}_{zu,2})$ &  & $.160$ & $.291$ & $.061$ &  & $.115$ & 
$.163$ & $.023$ &  & $.088$ & $.100$ & $.011$ &  & $.066$ & $.065$ & $.006$
\\ 
$\func{sd}(\hat{\sigma}_{zv,1})$ &  & $.327$ & $.558$ & $.125$ &  & $.259$ & 
$.329$ & $.053$ &  & $.184$ & $.209$ & $.024$ &  & $.120$ & $.142$ & $.011$
\\ 
$\func{sd}(\hat{\sigma}_{zv,2})$ &  & $.295$ & $.420$ & $.084$ &  & $.191$ & 
$.245$ & $.033$ &  & $.126$ & $.153$ & $.020$ &  & $.081$ & $.102$ & $.007$
\\ 
$\func{sd}(\hat{\Sigma}_{uv,11})$ &  & $.411$ & $.490$ & $.125$ &  & $.310$
& $.289$ & $.051$ &  & $.161$ & $.177$ & $.020$ &  & $.124$ & $.121$ & $.012$
\\ 
$\func{sd}(\hat{\Sigma}_{uv,21})$ &  & $.398$ & $.377$ & $.083$ &  & $.311$
& $.202$ & $.035$ &  & $.181$ & $.124$ & $.011$ &  & $.151$ & $.081$ & $.007$
\\ 
$\func{sd}(\hat{\Sigma}_{uv,12})$ &  & $.305$ & $.333$ & $.061$ &  & $.192$
& $.189$ & $.024$ &  & $.117$ & $.119$ & $.012$ &  & $.084$ & $.079$ & $.006$
\\ 
$\func{sd}(\hat{\Sigma}_{uv,22})$ &  & $.254$ & $.299$ & $.057$ &  & $.153$
& $.166$ & $.022$ &  & $.109$ & $.104$ & $.011$ &  & $.070$ & $.069$ & $.005$
\\ \hline
\end{tabular}

%TCIMACRO{\TeXButton{End sideways}{\end{sideways}}}%
%BeginExpansion
\end{sideways}%
%EndExpansion
\caption{Simulation Results. Comparison of true and asymptotic standard deviations of the parameter estimators 
($\times 10$). 
For the asymptotic standard deviations, mean and standard deviations are reported. Results for simulations with 
$\tau = \log 10$.}\label{tab:Sim_asd_10}%
%TCIMACRO{\TeXButton{E}{\end{table}}}%
%BeginExpansion
\end{table}%
%EndExpansion

%TCIMACRO{\TeXButton{B}{\begin{table}[tbp] \centering}}%
%BeginExpansion
\begin{table}[tbp] \centering%
%EndExpansion

%TCIMACRO{\TeXButton{Begin sideways}{\begin{sideways}}}%
%BeginExpansion
\begin{sideways}%
%EndExpansion

\begin{tabular}{ccccccccccccccccc}
\hline
&  & \multicolumn{15}{c}{$n$} \\ \cline{3-17}
&  & \multicolumn{3}{c}{$50$} &  & \multicolumn{3}{c}{$100$} &  & 
\multicolumn{3}{c}{$200$} &  & \multicolumn{3}{c}{$400$} \\ 
\cline{3-5}\cline{7-9}\cline{11-13}\cline{15-17}
Parameter &  & True & Mean & Sd &  & True & Mean & Sd &  & True & Mean & Sd
&  & True & Mean & Sd \\ \hline
$\func{sd}(\hat{\sigma}_{zu,1})$ &  & $.144$ & $.207$ & $.045$ &  & $.095$ & 
$.127$ & $.018$ &  & $.077$ & $.086$ & $.009$ &  & $.056$ & $.058$ & $.005$
\\ 
$\func{sd}(\hat{\sigma}_{zu,2})$ &  & $.118$ & $.151$ & $.034$ &  & $.081$ & 
$.092$ & $.014$ &  & $.058$ & $.061$ & $.006$ &  & $.040$ & $.041$ & $.003$
\\ 
$\func{sd}(\hat{\sigma}_{zv,1})$ &  & $.332$ & $.401$ & $.091$ &  & $.200$ & 
$.252$ & $.037$ &  & $.156$ & $.169$ & $.019$ &  & $.110$ & $.113$ & $.008$
\\ 
$\func{sd}(\hat{\sigma}_{zv,2})$ &  & $.217$ & $.269$ & $.056$ &  & $.141$ & 
$.169$ & $.022$ &  & $.092$ & $.111$ & $.012$ &  & $.071$ & $.076$ & $.006$
\\ 
$\func{sd}(\hat{\Sigma}_{uv,11})$ &  & $.322$ & $.373$ & $.099$ &  & $.209$
& $.223$ & $.041$ &  & $.147$ & $.151$ & $.020$ &  & $.096$ & $.101$ & $.010$
\\ 
$\func{sd}(\hat{\Sigma}_{uv,21})$ &  & $.240$ & $.200$ & $.037$ &  & $.169$
& $.121$ & $.018$ &  & $.108$ & $.079$ & $.009$ &  & $.078$ & $.054$ & $.004$
\\ 
$\func{sd}(\hat{\Sigma}_{uv,12})$ &  & $.139$ & $.185$ & $.041$ &  & $.125$
& $.110$ & $.013$ &  & $.082$ & $.073$ & $.006$ &  & $.057$ & $.050$ & $.004$
\\ 
$\func{sd}(\hat{\Sigma}_{uv,22})$ &  & $.130$ & $.162$ & $.033$ &  & $.081$
& $.100$ & $.015$ &  & $.066$ & $.067$ & $.007$ &  & $.041$ & $.045$ & $.003$
\\ \hline
\end{tabular}

%TCIMACRO{\TeXButton{End sideways}{\end{sideways}}}%
%BeginExpansion
\end{sideways}%
%EndExpansion
\caption{Simulation Results. Comparison of true and asymptotic standard deviations of the parameter estimators 
($\times 10$). 
For the asymptotic standard deviations, mean and standard deviations are reported. Results for simulations with 
$\tau = \log 30$.}\label{tab:Sim_asd_30}%
%TCIMACRO{\TeXButton{E}{\end{table}}}%
%BeginExpansion
\end{table}%
%EndExpansion

Tables \ref{tab:Sim_asd_10} and \ref{tab:Sim_asd_30} compare the true
finite-sample standard deviations of the estimators with their mean
asymptotic approximations. We use the `reduced'\ asymptotics mentioned at
the end of Section \ref{sec:Asymptotics}, treating the functional parameters
as if they were fixed and known. The dimension of the full $\mathbf{\theta }$
is 131, whereas the dimension of the reduced $\mathbf{\tilde{\theta}}$ is
14, so it is clear that only the `reduced' asymptotics is practical for our
sample sizes. Tables \ref{tab:Sim_asd_10} and \ref{tab:Sim_asd_30} show that
the true standard deviations of the estimators are accurately estimated,
especially for $n\geq 100$. Even for $n=50$, where the approximation is not
as good for some parameters, the asymptotic standard deviations tend to
overestimate the true standard deviations, which, for inferential purposes,
is better than underestimating them. For $n\geq 200$ the approximation is
extremely accurate for most parameters, even under the sparse scenario $\tau
=\log 10$. The accuracy of the approximation does not change much with $\tau 
$.

\section{Application: Chicago's Divvy bike sharing system\label{sec:Example}}

As mentioned in the Introduction, in this section we analyze bike trips that
took place between April 1 and November 31 of 2016 in Chicago's Divvy
system. Specifically, we analyze trips originating at station 166, located
at the intersection of Wrightwood and Ashland avenues. For each bike trip we
observe the time $t$ when the bike was checked out and its spatial
destination $\mathbf{s}$, so we can see $(t,\mathbf{s})$ as an observation
from a spatio-temporal process. Strictly speaking, $\mathbf{s}$ is a
discrete variable that can only take values on the lattice of 458 stations,
but this grid is dense enough that for practical purposes we can consider $%
\mathbf{s}$ as continuous.

For estimation of the temporal functional parameters we used cubic B-splines
with ten equally spaced knots in $B_{t}=[0,24]$, so the family $\mathcal{B}%
_{t}$ had dimension $q_{1}=14$. The spatial domain $B_{s}$ is more
irregular. Since all trips from this station have destinations within the
rectangle $[-87.840,-87.530]\times \lbrack 41.800,42.030]$, we took as $%
B_{s} $ the sector of the city included in this rectangle, which is
basically the northern half of the city. As basis family for the spatial
functional parameters we used renormalized Gaussian kernels with 43 equally
spaced centroids (we created a grid of 100 equally spaced points in the
rectangle $[-87.840,-87.530]\times \lbrack 41.800,42.030]$, and 43 of those
ended up within the city boundaries). Then the family $\mathcal{B}_{s}$ has
dimension $q_{2}=43$. As smoothing parameters we took all $\xi $s equal to $%
10^{-5}$, which provide smooth estimators while retaining a reasonable level
of local detail.

We tried different combinations of numbers of components $(p_{1},p_{2})$: $%
(1,1)$, $(2,2)$, $(3,2)$, $(3,3)$ and $(4,4)$. For each model we computed
five-fold cross-validated mean log-likelihoods, obtaining $40.61$, $41.23$, $%
41.34$, $41.46$ and $41.50$, respectively. A scree-plot shows that there is
a big improvement from the $(1,1)$-model to the $(2,2)$-model, but
practically no improvement from the $(3,3)$-model to the $(4,4)$-model. For
the $(3,3)$-model the relative contribution of the variances of the spatial
components are 82\%, 16\% and 2\%, respectively, so the last component is
rather superfluous. For this reason we opted for the $(2,2)$-model, where
the relative variance proportions for the temporal components are 67\% and
33\%, and for the spatial components 75\% and 25\%, respectively.

\FRAME{ftbpFU}{6.058in}{2.623in}{0pt}{\Qcb{Divvy Data Analysis. Effect of
the temporal components on the baseline intensity. (a) First components, (b)
second component. Plot shows baseline intensities (solid line), plus (dashed
line) and minus (dotted line) a multiple of the component.}}{\Qlb{%
fig:temp_pcs}}{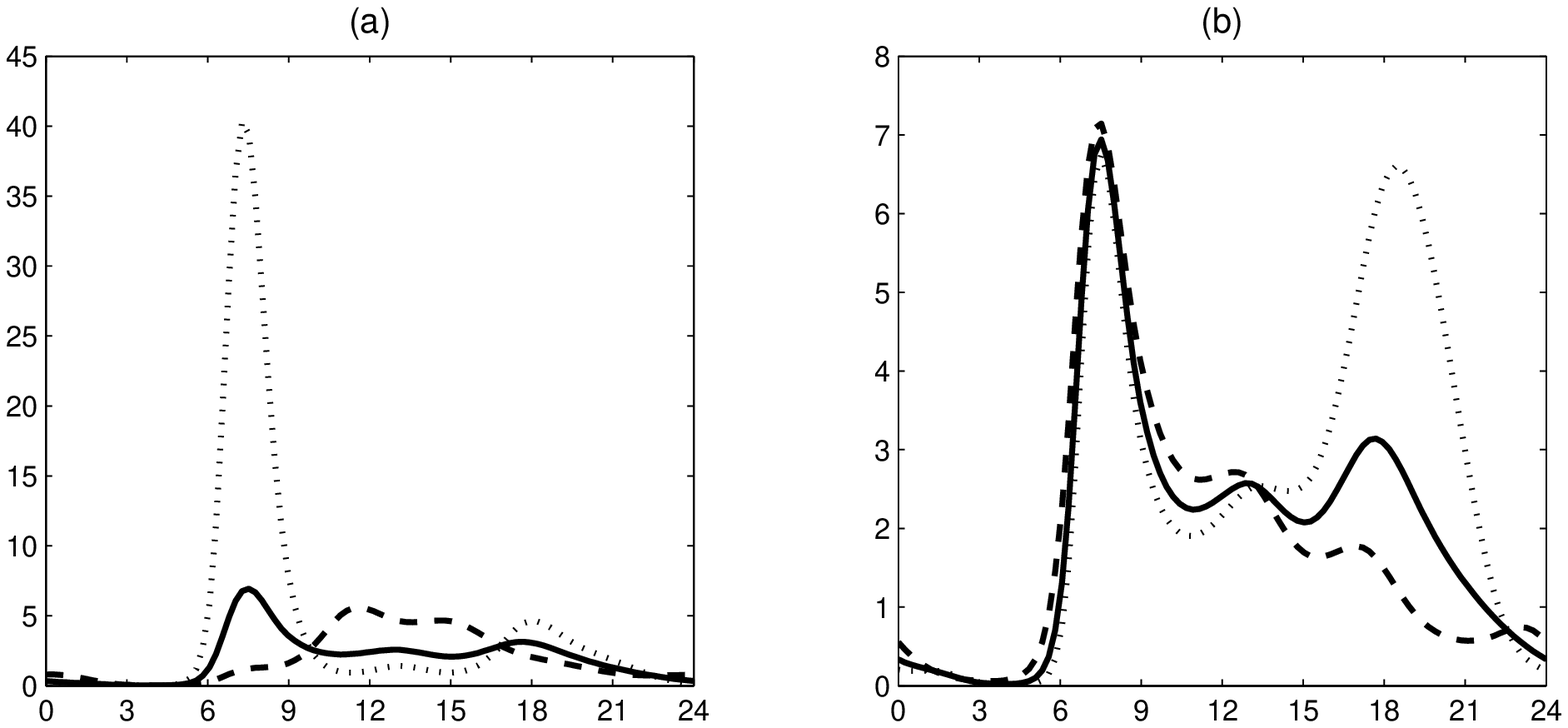}{\special{language "Scientific Word";type
"GRAPHIC";maintain-aspect-ratio TRUE;display "USEDEF";valid_file "F";width
6.058in;height 2.623in;depth 0pt;original-width 9.2907in;original-height
3.6919in;cropleft "0.0768";croptop "1";cropright "1";cropbottom "0";filename
'temp_pcs.eps';file-properties "XNPEU";}}

To interpret the temporal components we plotted $\exp \{\hat{\mu}(t)\}$
versus $\exp \{\hat{\mu}(t)\pm c\hat{\phi}_{k}(t)\}$ for each $\hat{\phi}%
_{k} $, where $c$ is a constant chosen for convenient visualization. Figure %
\ref{fig:temp_pcs}(a) shows that a negative score on the first component
corresponds to a sharp morning peak around 7 am, while a positive score is
associated with the absence of a morning spike and a higher bike demand in
the early afternoon. This component accounts for the difference between
weekday and weekend patterns of demand. This is corroborated by a time
series plot of the scores $\hat{u}_{i1}$, shown in the Supplementary
Material, which is strongly weekly periodic with peaks occurring on Sundays
and troughs on Thursdays or Wednesdays. Figure \ref{fig:temp_pcs}(b) shows
that a negative score on the second component is associated with higher bike
demand in the evening, around 6 pm, while a positive score is associated
with lower demand at this time. The time series plot of the scores $\hat{u}%
_{i2}$ in the Supplementary Material shows a clear seasonal trend, with a
minimum at the summer months. So this component is associated with a
seasonal pattern of demand.

\FRAME{ftbpFU}{5.9845in}{2.623in}{0pt}{\Qcb{Divvy Data Analysis. Effect of
the first spatial component on the baseline intensity. Plots show baseline
intensity minus [(a)] and plus [(b)] a multiple of the component.}}{\Qlb{%
fig:sp_pc_1}}{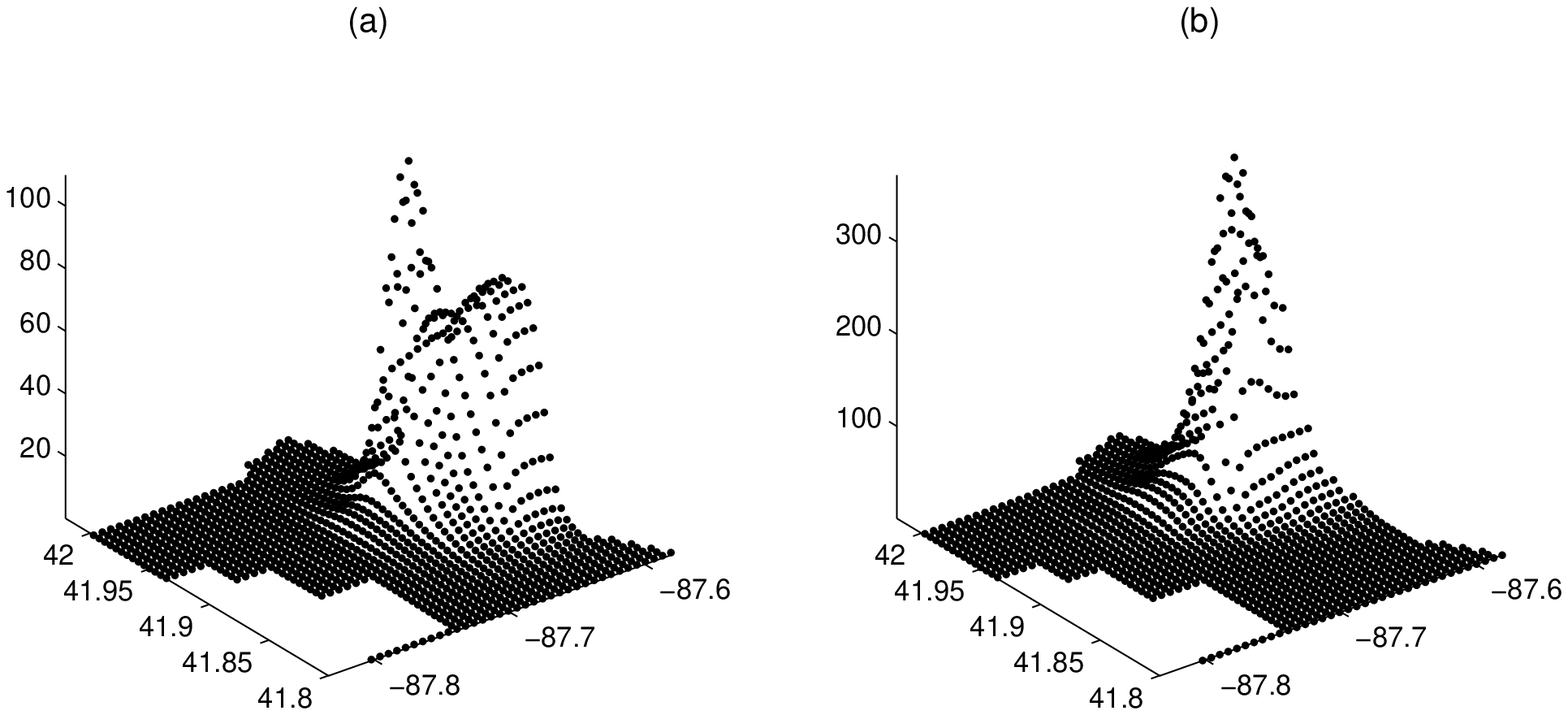}{\special{language "Scientific Word";type
"GRAPHIC";maintain-aspect-ratio TRUE;display "ICON";valid_file "F";width
5.9845in;height 2.623in;depth 0pt;original-width 9.2907in;original-height
3.7343in;cropleft "0.0777";croptop "1";cropright "1";cropbottom "0";filename
'sp_pc_1.eps';file-properties "XNPEU";}}

Spatial components are harder to interpret from static plots, so we provide
three-dimensional surface plots here and color contour plots in the
Supplementary Material. Figure \ref{fig:sp_pc_1} corresponds to the first
component. We see that a positive score corresponds to a sharp peak at the
bike station itself, meaning that most trips are short and local those days.
A negative score is associated with a higher proportion of trips downtown. A
time series plot of the $\hat{v}_{i1}$s, shown in the Supplementary
Material, is strongly weekly periodical, indicating that this component is
strongly associated with weekday versus weekend patterns of usage. For the
second spatial component, Figure \ref{fig:sp_pc_2} shows that positive
scores are associated with days when most trips stay within the neighborhood
or downtown, whereas negative scores correspond to days with a higher
proportion of faraway trips.

\FRAME{ftbpFU}{5.8427in}{2.6221in}{0pt}{\Qcb{Divvy Data Analysis. Effect of
the second spatial component on the baseline intensity. Plots show baseline
intensity minus [(a)] and plus [(b)] a multiple of the component.}}{\Qlb{%
fig:sp_pc_2}}{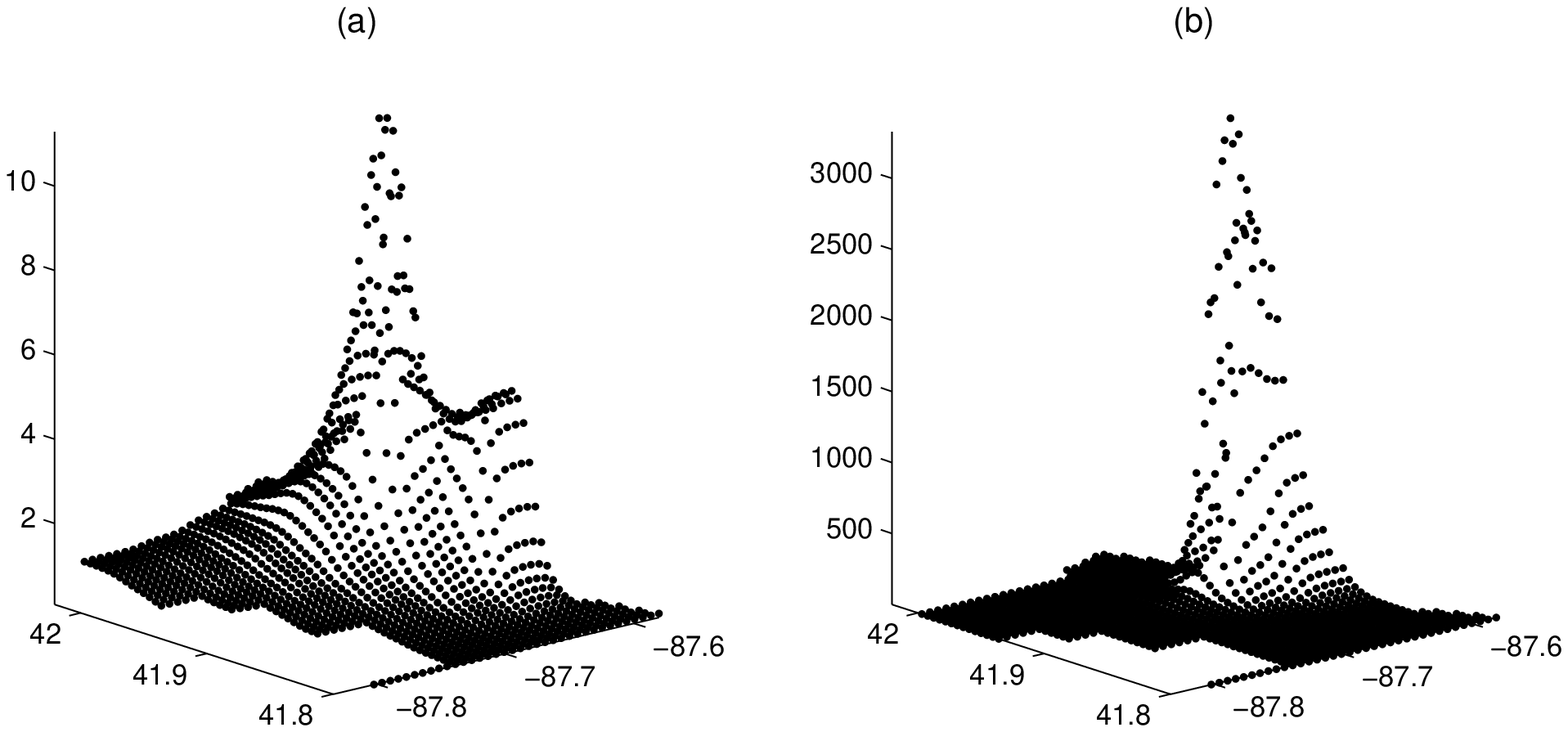}{\special{language "Scientific Word";type
"GRAPHIC";maintain-aspect-ratio TRUE;display "ICON";valid_file "F";width
5.8427in;height 2.6221in;depth 0pt;original-width 9.2907in;original-height
3.8164in;cropleft "0.0794";croptop "1";cropright "1";cropbottom "0";filename
'sp_pc_2.eps';file-properties "XNPEU";}}

The estimated cross-correlations between temporal and spatial component
scores are: $\limfunc{corr}(U_{1},V_{1})=.90$, $\limfunc{corr}%
(U_{1},V_{2})=.32$, $\limfunc{corr}(U_{2},V_{1})=-.12$ and $\limfunc{corr}%
(U_{2},V_{2})=.10$. The asymptotic standard deviations of these estimators,
derived from the results in Section \ref{sec:Asymptotics} using the Delta
Method, are $.10$, $.12$, $.19$ and $.19$, respectively. Therefore only $%
\limfunc{corr}(U_{1},V_{1})$ and $\limfunc{corr}(U_{1},V_{2})$ are
statistically significant. The high correlation between $U_{1}$ and $V_{1}$
is not surprising and easy to interpret: on weekdays, there is a higher
proportion of bike trips early in the morning with a downtown destination,
suggesting that people use bikes for work commute; whereas on weekends, most
bike trips take place in the afternoon and tend to stay in the neighborhood.

\section*{Acknowledgement}

This research was partly supported by US National Science Foundation grant
DMS 1505780.

\section*{References}

\begin{description}
\item Ash, R.B. and Gardner, M.F. (1975). \emph{Topics in stochastic
processes}. Academic Press, New York.

\item Baddeley, A. (2007). Spatial point processes and their applications.
In \emph{Stochastic Geometry}, Lecture Notes in Mathematics 1892, pp.~1--75.
Springer, New York.

\item Baddeley, A.J., Moyeed, R.A., Howard, C.V., and Boyde, A. (1993).
Analysis of a three-dimensional point pattern with replication. \emph{%
Applied Statistics} \textbf{42 }641--668.

\item Bell, M.L., and Grunwald, G.K. (2004). Mixed models for the analysis
of replicated spatial point patterns. \emph{Biostatistics }\textbf{5 }%
633--648.

\item Bouzas, P.R., Valderrama, M., Aguilera, A.M., and Ruiz-Fuentes, N.
(2006). Modelling the mean of a doubly stochastic Poisson process by
functional data analysis. \emph{Computational Statistics and Data Analysis} 
\textbf{50} 2655--2667.

\item Bouzas, P.R., Ruiz-Fuentes, N., and Oca\~{n}a, F.M. (2007). Functional
approach to the random mean of a compound Cox process. \emph{Computational
Statistics }\textbf{22} 467--479.

\item Brown, E.N., Kass, R.E. and Mitra, P.P. (2004). Multiple neural spike
train data analysis: state-of-the-art and future challenges. \emph{Nature
Neuroscience }\textbf{7} 456--461.

\item Cox, D.R., and Isham, V. (1980). \emph{Point Processes.} Chapman and
Hall/CRC, Boca Raton.

\item Dempster, A.P., Laird, N.M., and Rubin, D.B. (1977). Maximum
likelihood from incomplete data via the EM algorithm. \emph{Journal of the
Royal Statistical Society Series B} \textbf{39} 1--38.

\item Diggle, P.J. (2013). \emph{Statistical Analysis of Spatial and
Spatio-Temporal Point Patterns, Third Edition.} Chapman and Hall/CRC, Boca
Raton.

\item Diggle, P.J., Lange, N., and Bene\v{s}, F.M. (1991). Analysis of
variance for replicated spatial point patterns in clinical neuroanatomy. 
\emph{Journal of the American Statistical Association }\textbf{86} 618--625.

\item Diggle, P.J., Mateau, J., and Clough, H.E. (2000). A comparison
between parametric and nonparametric approaches to the analysis of
replicated spatial point patterns. \emph{Advances in Applied Probability} 
\textbf{32 }331--343.

\item Diggle, P.J., Eglen, S.J., and Troy, J.B. (2006). Modeling the
bivariate spatial distribution of amacrine cells. In \emph{Case Studies in
Spatial Point Process Modeling}, eds.~A. Baddeley et al., New York:
Springer, pp.~215--233.

\item Fern\'{a}ndez-Alcal\'{a}, R.M., Navarro-Moreno, J., and Ruiz-Molina,
J.C. (2012). On the estimation problem for the intensity of a DSMPP. \emph{%
Methodology and Computing in Applied Probability }\textbf{14} 5--16.

\item Gervini, D. (2016). Independent component models for replicated point
processes. \emph{Spatial Statistics }\textbf{18} 474--488.

\item Gervini, D. and Baur, T.J. (2017). Joint models for grid point and
response processes in longitudinal and functional data. To appear in \emph{%
Statistica Sinica} (currently \emph{ArXiv} \textbf{1705.06259}).

\item Gervini, D. and Khanal, M. (2019). Exploring patterns of demand in
bike sharing systems via replicated point process models. \emph{Journal of
the Royal Statistical Society Series C: Applied Statistics} \textbf{68}
585--602.

\item Geyer, C.J. (1994). On the asymptotics of constrained M-estimation. 
\emph{The Annals of Statistics }\textbf{22} 1993--2010.

\item Hastie, T., Tibshirani, R., and Friedman, J. (2009). \emph{The
Elements of Statistical Learning. Data Mining, Inference, and Prediction.
Second Edition. }Springer, New York.

\item Knight, K., and Fu, W. (2000). Asymptotics for lasso-type estimators. 
\emph{The Annals of Statistics }\textbf{28} 1356--1378.

\item Landau, S., Rabe-Hesketh, S., and Everall, I.P. (2004). Nonparametric
one-way analysis of variance of replicated bivariate spatial point patterns. 
\emph{Biometrical Journal }\textbf{46 }19--34.

\item Li, Y., and Guan, Y. (2014). Functional principal component analysis
of spatiotemporal point processes with applications in disease surveillance. 
\emph{Journal of the American Statistical Association }\textbf{109}
1205--1215.

\item M\o ller, J., and Waagepetersen, R.P. (2004). \emph{Statistical
Inference and Simulation for Spatial Point Processes}. Chapman and Hall/CRC,
Boca Raton.

\item Nair, R., and Miller-Hooks, E. (2011). Fleet management for vehicle
sharing operations. \emph{Transportation Science }\textbf{45 }524--540.

\item Pawlas, Z. (2011). Estimation of summary characteristics from
replicated spatial point processes. \emph{Kybernetika} \textbf{47} 880--892.

\item Pollard, D. (1984). \emph{Convergence of Stochastic Processes. }%
Springer, New York.

\item Ramsay, J.O., and Silverman, B.W. (2005). \emph{Functional Data
Analysis (second edition)}. Springer, New York.

\item Rockafellar, R.T., and Wets, R.J. (1998). \emph{Variational Analysis.}
Springer, New York.

\item Ruppert, D. (2002). Selecting the number of knots for penalized
splines. \emph{Journal of Computational and Graphical Statistics} \textbf{11}
735--757.

\item Seber, G.A.F. (2004). \emph{Multivariate Observations.} Wiley, New
York.

\item Shaheen, S., Guzman, S., and Zhang, H. (2010). Bike sharing in Europe,
the Americas and Asia: Past, present and future. \emph{Transportation
Research Record: Journal of the Transportation Research Board }\textbf{2143}
159--167.

\item Shirota, S., and Gelfand, A.E. (2017). Space and circular time log
Gaussian Cox processes with application to crime event data. \emph{The
Annals of Applied Statistics }\textbf{11} 481--503.

\item Streit, R.L. (2010). \emph{Poisson Point Processes: Imaging, Tracking,
and Sensing.} Springer, New York.

\item Van der Vaart, A. (2000). \emph{Asymptotic Statistics}. Cambridge
University Press, Cambridge, UK.

\item Waagepetersen, R., Guan, Y., Jalilian, A., and Mateu, J. (2016).
Analysis of multispecies point patterns by using multivariate log-Gaussian
Cox processes. \emph{Journal of the Royal Statistical Society Series C:
Applied Statistics} \textbf{65} 77--96.

\item Wager, C.G., Coull, B.A., and Lange, N. (2004). Modelling spatial
intensity for replicated inhomogeneous point patterns in brain imaging. 
\emph{Journal of the Royal Statistical Society Series B} \textbf{66 }%
429--446.

\item Wu, S., M\"{u}ller, H.-G., and Zhang, Z. (2013). Functional data
analysis for point processes with rare events. \emph{Statistica Sinica }%
\textbf{23} 1--23.

\item Xun, X., Cao, J., Mallick, B., Maity, A., and Carroll, R.J. (2013).
Parameter estimation of partial differential equations. \emph{Journal of the
American Statistical Association} \textbf{108} 1009--1020.

\item Yu, Y., and Ruppert, D. (2002). Penalized spline estimation for
partially linear single-index models. \emph{Journal of the American
Statistical Association} \textbf{97} 1042--1054.
\end{description}

\end{document}